%% file: main.tex
\documentclass[11pt,letterpaper]{article}

\usepackage[margin=1in]{geometry}
\usepackage[utf8]{inputenc}
\usepackage[T1]{fontenc}
\usepackage{microtype}
\usepackage{lmodern}

\usepackage{amsmath}
\usepackage{amssymb}
\usepackage{mathtools}

\usepackage{booktabs}
\usepackage{multirow}
\usepackage{graphicx}
\usepackage{listings}
\usepackage{xcolor}

\usepackage[numbers,sort&compress]{natbib}
\usepackage{url}
\usepackage{hyperref}
\hypersetup{
  colorlinks = true,
  linkcolor  = black,
  urlcolor   = black,
  citecolor  = black,
}

\lstdefinestyle{pseudocode}{
  basicstyle    = \ttfamily\small,
  keywordstyle  = \bfseries,
  commentstyle  = \itshape\color{gray},
  numberstyle   = \tiny\color{gray},
  showstringspaces = false,
  breaklines    = true,
  frame         = single,
  framesep      = 4pt,
  rulecolor     = \color{gray!30},
  morekeywords  = {on, while, for, in, if, return, break, continue},
}
\title{A Biophysically-Inspired Feedback Controller \\
       for Multi-Class Cache Fairness}
\author{
  Matt. R. Flax\\
  \texttt{flatmax@flatmax.com}
}
\date{}

\begin{document}
\maketitle

\begin{abstract}
Cache replacement under multi-tenant LLM-serving conditions is a
multi-class problem: short, high-reuse system prompts; long,
moderate-reuse user documents; medium-length code context; and
bursty conversation history share a single eviction pool. Under
skewed multi-class arrivals, conventional flat-LRU policies expose
the worst-served-class miss ratio ($m_{\max}$) only as a fixed point.
We introduce a class of cache-replacement policies parameterised by a
per-class flux formula, where three structural commitments --- a
single global token-mass imbalance signal, $K$ parallel rectified
per-class promotion accumulators, and an age-ordered eviction
backstop --- produce emergent multi-class fairness. We instantiate
this class with a linear V-coupled rectified flux and a
Goldman-Hodgkin-Katz extension whose $V \to 0$ limit is exactly
the linear form. Across four skew levels on synthetic multi-class
workloads, the policy class closes 27--72\,\% of the
LRU$\to$Belady gap on $m_{\max}$, with linear and GHK
interchangeable on the headline objective within search variance. The fairness/throughput tradeoff is
exposed as a tunable knob on a single hyperparameter axis. We
position this against the LeCaR feedback-controller lineage and the
formal-control-theory cache-decay lineage as a novel combination of
known ingredients. Code and reproduction scripts:
\url{https://github.com/flatmax/membrane.cache}.
\end{abstract}

\input{sections/01_introduction.tex}
\input{sections/02_background.tex}
\input{sections/03_policy.tex}
\input{sections/04_implementation.tex}
\input{sections/05_methodology.tex}
\input{sections/06_results.tex}
\input{sections/07_discussion.tex}
\input{sections/08_related_work.tex}
\input{sections/09_limitations.tex}

\section*{Acknowledgements}
\label{sec:acks}

\paragraph{AI tool disclosure.}
Claude (Anthropic) was used as a writing and coding assistant during
the preparation of this manuscript: drafting and revising prose
across all sections, generating \LaTeX{} tables and figures from CSV
experiment outputs, and assisting with the implementation and
analysis scripts under \texttt{scripts/}. The author reviewed,
verified, and accepts full responsibility for all content, including
all numerical results, citations, and claims.

\bibliographystyle{plainnat}
\bibliography{refs}

\end{document}

%% file: sections/01_introduction.tex
\section{Introduction}
\label{sec:intro}

\subsection{The multi-class problem}

Cache replacement under multi-tenant LLM-serving conditions is a
\emph{multi-class} problem. A production LLM-serving stack
\citep{kwon2023vllm,zheng2024sglang} sees at least four
qualitatively different content classes flowing through its prefix
cache: short, high-reuse system prompts; long, moderate-reuse user
documents; medium-length code context; and bursty, low-reuse
conversation history. These classes differ in arrival rate, token
mass, and locality structure by an order of magnitude or more, and
they share a single eviction pool.

Conventional flat-LRU policies treat the pool as single-class. Under
skewed multi-class arrivals --- the empirical regime of multi-tenant
serving --- this means one class's burst can starve the others: a
long docs-class request stream evicts the system-prompt class
faster than it can be re-warmed, and the system-prompt miss ratio
climbs while the aggregate miss ratio looks healthy. The relevant
fairness objective is \emph{not} the aggregate, but the worst-served
class --- the \textbf{max-class miss ratio} $m_{\max} = \max_k m_k$,
the metric that drives head-of-line blocking on the slowest tenant.

The trouble with $m_{\max}$ as a tuning target is that single-pool
policies expose it as a \emph{fixed point}. LRU gives one $m_{\max}$
value; ARC \citep{megiddo2003arc} gives another; TinyLFU gives a
third. There is no knob to trade worst-class fairness against
aggregate throughput without forking the policy. This is the gap
the present work closes.

\subsection{What we propose}

We derive and evaluate a class of cache-replacement policies
parameterised by a per-class flux formula, where three structural
commitments --- (i) a single global token-mass imbalance signal
driving (ii) $K$ parallel rectified per-class promotion accumulators,
with eviction handled by (iii) an age-ordered backstop --- produce
\emph{emergent} multi-class fairness. The fairness/throughput
tradeoff is exposed as a tunable knob on a single hyperparameter
axis ($V_T$, the soft-knee scale), with one endpoint matching LRU's
byte miss ratio to four decimal places and the other closing 72\,\%
of the LRU$\to$Belady gap on $m_{\max}$ at the moderate-skew
operating point.

We instantiate this class with two formulas:

\begin{itemize}
  \item a \textbf{linear V-coupled rectified flux}
        (\S\ref{sec:policy:linear}),
        $\Phi_k = \max(0,\; P \cdot V \cdot (c_{l,k} - c_{u,k}) / V_T)$,
        derived as the minimal controller satisfying the three
        structural commitments;
  \item a \textbf{Goldman-Hodgkin-Katz extension}
        (\S\ref{sec:policy:ghk}) whose $V \to 0$ Taylor limit
        \emph{is} the linear form, with an exponential weighting
        that sharpens response under zero-concentration conditions.
\end{itemize}

The contribution of this paper is the \emph{class}, not the formula.
\S\ref{sec:results} shows that linear and GHK tie exactly at the
moderate-skew operating point and differ by only 1--4 percentage
points of $m_{\max}$ at the others, with neither dominating across
the four skew levels. The signal coupling and the rectification
clamp are what does the work; the specific functional form is a
secondary tuning surface that helps on harder objectives (Jain's
index) but is interchangeable on the headline within search
variance.

\subsection{Pre-empting the diffusion-cache naming collision}

A growing body of LLM-systems work uses \emph{diffusion cache} to
mean ``a cache for the intermediate states of a diffusion
\emph{generative} model'' --- Sparse-dLLM, NIRVANA, SenCache, and
SeaCache all fall in this lineage. They use ordinary LRU/LFU on a
niche workload; they do not use the physics of diffusion as a
\emph{control law}.

Our usage is orthogonal. We use \emph{electrodiffusion physics}
--- specifically the Goldman-Hodgkin-Katz flux equation
\citep{goldman1943potential,hodgkin1949effect,hille2001ion} and its
V-coupled linear limit --- as the mechanism that drives a
cache-replacement controller. This is a controller-derivation
choice, not a workload choice. The two literatures share no overlap
of techniques; we flag the naming collision here so that later
sections can use the term \emph{flux} without ambiguity.

\subsection{Position vs. prior art}

Our policy is best characterised as a \emph{biophysically-inspired
feedback controller for multi-class memory hierarchies}. The
closest existing precedents fall in three lineages, and we depart
from each:

\begin{itemize}
  \item \textbf{Feedback-driven cache controllers} (lineage anchor:
        LeCaR \citep{vietri2018lecar}). LeCaR weights pure LRU and
        pure LFU by a regret signal --- single-class, binary,
        history-driven. We are multi-class with a continuous global
        signal driving per-class accumulators.
  \item \textbf{Formal control theory in caches} (lineage anchor:
        Velusamy et al.\ \citep{velusamy2002imc}). Their integral
        controller adapts a cache-line \emph{deactivation timer}; we
        adapt a per-class \emph{promotion rate}.
  \item \textbf{Multi-tenant cache fairness} (FairRide
        \citep{pu2016fairride}, Cliffhanger
        \citep{cidon2016cliffhanger}, Memshare
        \citep{cidon2017memshare}, RobinHood
        \citep{berger2018robinhood}, Hyperbolic
        \citep{blankstein2017hyperbolic}, vLLM, SGLang). These
        either partition or schedule with priority; none expose a
        continuous tunable fairness/throughput tradeoff under a
        single global signal.
\end{itemize}

Detailed comparison is in \S\ref{sec:related}. We treat the GHK
biology origin as \emph{recognition} of a principled functional
form, not as a source of authority for it
(\S\ref{sec:policy:ghk}); the controller derives from
caching-first constraints alone.

\subsection{Contributions and roadmap}

This paper makes four contributions:

\begin{enumerate}
  \item A controller-derivation framework
        (\S\ref{sec:policy}) starting from three caching-first
        constraints --- global signal coupling, rectification, and
        graceful behaviour at zero concentration --- that admits a
        small family of multi-class promotion-rate controllers.
  \item A linear V-coupled rectified flux
        (\S\ref{sec:policy:linear}) as the minimal element of that
        family, and a GHK extension (\S\ref{sec:policy:ghk}) as the
        principled non-linear member.
  \item A libCacheSim plugin implementation
        (\S\ref{sec:impl}) with deterministic replays under fixed
        seeds, and a methodology (\S\ref{sec:method}) that holds
        Optuna training seeds disjoint from test seeds.
  \item An empirical evaluation (\S\ref{sec:results}) across four
        skew levels on synthetic multi-class workloads, showing
        that the policy class closes 27\,\%--72\,\% of the
        LRU$\to$Belady gap on $m_{\max}$, with linear and GHK
        interchangeable on the headline objective within search
        variance and GHK earning its keep on the Jain-tuned
        secondary objective.
\end{enumerate}

The remainder of the paper is organised as follows.
\S\ref{sec:background} sets out the prefix-cache cost model and the
four fairness metrics we report on. \S\ref{sec:policy} derives the
controller. \S\ref{sec:impl} covers the libCacheSim plugin
implementation. \S\ref{sec:method} describes the synthetic workload
generator and Optuna methodology. \S\ref{sec:results} reports
results. \S\ref{sec:discussion} discusses what the controller is
and is not good for. \S\ref{sec:related} covers related work in
detail. \S\ref{sec:limits} lists limitations and future work.

%% file: sections/02_background.tex
\section{Background and motivation}
\label{sec:background}

\subsection{The prefix-cache cost model}
\label{sec:bg:cost}

Modern multi-tenant LLM-serving stacks (vLLM PagedAttention
\citep{kwon2023vllm}, SGLang RadixAttention
\citep{zheng2024sglang}) keep a \textbf{prefix cache} of attention
key/value tensors keyed by the prefix of input tokens. A request
that shares a prefix with a previously cached request reuses the
cached tensors and skips the prefill cost on the shared portion.
The cache is sized in tokens (typically hundreds of GB on a
multi-GPU node) and managed by a replacement policy when capacity
is reached.

The unit of admission, eviction, and accounting is the \emph{paged}
KV block --- fixed-size pages holding the attention state for a
contiguous run of tokens. For the purposes of replacement-policy
design, each page is a cache line with two costs:

\begin{itemize}
  \item \textbf{Object cost}: a miss requires re-running the
        prefill kernel for that page's tokens, costing $O(n^2)$
        attention compute in the page size $n$.
  \item \textbf{Byte cost}: the page occupies a fixed number of
        bytes of HBM/DRAM, traded against other pages.
\end{itemize}

Aggregate \textbf{object miss ratio} $m_{\text{obj}}$ and
\textbf{byte miss ratio} $m_{\text{byte}}$ are the standard
flat-cache reporting metrics. They diverge whenever pages have
non-uniform byte size (long-context pages are larger), which is the
regime of interest here.

\subsection{Why multi-class fairness matters operationally}
\label{sec:bg:hol}

The four content classes that flow through a production prefix
cache --- system prompt, user documents, code context, conversation
history --- differ structurally:

\begin{itemize}
  \item \textbf{System prompt} is short, high-reuse, near-static.
        Misses cause \emph{every} request that uses the affected
        prompt to wait on prefill.
  \item \textbf{User documents} are long, moderate-reuse, often
        bursty (a document is loaded into context once per
        conversation and reused tens of turns).
  \item \textbf{Code context} sits between docs and history:
        medium-length, moderately reused within a coding session.
  \item \textbf{Conversation history} is bursty and
        low-reuse-after-completion: a session's history pages are
        hot during the session and cold immediately after.
\end{itemize}

The arrival-rate disparity is large --- system prompt and history
each typically supply ${\sim}5\,\%$ of the incoming page stream,
while docs and code each supply ${\sim}45\,\%$ under steady state.
A flat-LRU policy applied to the shared pool will, under a
sustained docs/code burst, evict system-prompt and history pages
faster than re-warming can keep up. The immediate operational
consequence is \textbf{head-of-line blocking on system-prompt
misses}: every request sharing the affected system prompt now waits
on a prefill the cache should have served. Aggregate miss ratio
looks healthy; the worst-served class is in trouble.

This is the multi-class fairness problem in concrete form. It is
not a problem of \emph{aggregate} cache behaviour; it is a problem
of the \emph{worst-case} class within the cache.

\subsection{The four fairness metrics we report}
\label{sec:bg:metrics}

There is no single agreed fairness metric for multi-class caches.
We report four metrics that have been used across the
cache-fairness literature (see \S\ref{sec:related}), each sensitive
to a different failure mode:

\begin{enumerate}
  \item \textbf{Max-class miss ratio} ($m_{\max} = \max_k m_k$).
        The strongest fairness statement: ``no class starves.''
        Sensitive to the worst-served class directly. This is the
        metric that tracks head-of-line blocking; we anchor the
        headline on it.
  \item \textbf{Per-class miss-ratio std-dev}
        ($\sigma_m = \mathrm{stdev}_k(m_k)$). The classical
        network-fairness flavour: how dispersed are the per-class
        outcomes? Lower = fairer.
  \item \textbf{Jain's fairness index}
        ($J = (\sum_k h_k)^2 / (K \sum_k h_k^2)$ over per-class
        hit rates $h_k$). Bounded in $(1/K, 1]$, with $1$ meaning
        perfectly equal hit rates. Familiar to systems reviewers.
  \item \textbf{Aggregate miss ratio}
        ($m_{\text{byte}}, m_{\text{obj}}$). The throughput axis.
        A fair-but-bad policy is not useful; we report the
        aggregate to confirm the policy class does not trade away
        substantial throughput for fairness.
\end{enumerate}

The first three measure fairness; the fourth measures throughput.
The contribution of \S\ref{sec:results} is that the
fairness/throughput tradeoff is \emph{exposed and tunable} on the
proposed policy class --- a tuning sweep that moves $m_{\max}$ from
$0.61$ (LRU) toward $0.45$ ($72\,\%$ of the LRU$\to$Belady gap
closed) costs aggregate $m_{\text{byte}}$ in a measurable,
predictable way, where every baseline policy gives a fixed point on
this Pareto.

\subsection{The headline metric}
\label{sec:bg:headline}

Of the four, $m_{\max}$ is the operationally load-bearing metric:
it is the one that maps directly to user-visible tail latency under
multi-tenant serving. Jain and std-dev are \emph{symmetric}
fairness measures --- they punish a class doing too well as much
as a class doing too poorly; neither is the right \emph{objective}
for a cache replacement controller, even though both are
reasonable to \emph{report}.

We anchor \S\ref{sec:results} on $m_{\max}$ for this reason. The
other three metrics are reported alongside as a fairness profile,
and the methodological observation in \S\ref{sec:discussion}
(``optimise on $m_{\max}$, report on Jain'') falls out of this
distinction.

%% file: sections/03_policy.tex
\section{Policy derivation}
\label{sec:policy}

\paragraph{Notation.}
We write $K$ for the number of content classes; $c_{l,k}$ and
$c_{u,k}$ for the count of class-$k$ files in the lower
(evictable) and upper (active) cache regions; $T_{l,k}$ and
$T_{u,k}$ for the token mass of class $k$ on each side; $P$ for a
single per-class promotion rate constant; $V_T$ for a thermal-
voltage analogue setting the soft-knee scale; and $\Phi_k$ for
the (signed) per-class promotion rate.

\subsection{Constraints}
\label{sec:policy:constraints}

We set out three caching-first design constraints before
introducing any formula. The contribution of this section is that
all three are individually defensible without invoking biology, and
their conjunction admits a small family of controllers --- not a
unique one.

\paragraph{(C1) Global cache pressure drives per-class promotion.}
Per-class admission control is widely treated as a partitioning
decision: each class is given a fixed cache budget that it manages
independently (e.g.\ utility-based cache partitioning, dedicated
LRU/LFU per tenant). We make the opposite choice. There is one
global signal --- the cache-wide imbalance between upper-tier and
lower-tier token mass

\begin{equation}
V \;=\; \sum_{k=1}^{K} \bigl(T_{l,k} - T_{u,k}\bigr)
\label{eq:V}
\end{equation}

--- and every class's promotion is driven by it. Tokens are the
universal mass carried by every file regardless of class; they sum
into one scalar. This is the structural commitment that produces
emergent fairness: under-represented classes share the same
driving signal as over-represented ones, but their \emph{response}
to that signal differs because of (C3).

The alternative --- local per-class queue pressure (the WFQ / HTB
lineage) --- gives every class a guaranteed share regardless of
what other classes are doing. We choose global because it admits a
\emph{tunable} fairness/throughput tradeoff
(\S\ref{sec:results:sensitivity}) where the local-signal lineage
exposes only a single point per parameter choice. The architectural
lineage and the closer AQM PI-controller cousins
\citep{rfc8033} are \S\ref{sec:related}.

\paragraph{(C2) Promotion must rectify.}
Cache lines in the lower (evictable) region are demoted by an
age-ordered backstop, not by negative promotion pressure from the
controller. The controller's output is therefore non-negative:
$\Phi_k \geq 0$. Negative pressure would require the controller to
choose \emph{which} class to demote under shared eviction, and we
delegate that to the backstop deliberately --- eviction-class
selection from imbalance signals admits search-space pathologies
that appear in our experiments as Jain-objective tunes that
equalise everyone toward worse rather than raising minorities
toward better (\S\ref{sec:results:linear}).

\paragraph{(C3) The controller must admit a non-linear extension
near zero concentration.}
A class can have zero representation in either region. The
controller must behave well at the boundary $c_{u,k} \to 0$ and
$c_{l,k} \to 0$, and it should expose a tuning surface that admits
non-linear weighting of the over-represented side without numerical
pathology. Constraint (C3) is the weakest of the three --- it
admits any controller whose linear and non-linear forms differ only
by a multiplier that is $1 + O(V)$. We use it to motivate the GHK
form in \S\ref{sec:policy:ghk} as a principled choice from a small
space of candidates.

\subsection{The linear V-coupled rectified flux}
\label{sec:policy:linear}

The minimal controller satisfying (C1)+(C2) is

\begin{equation}
\Phi_k \;=\; \max\!\left(0,\; \frac{P \cdot V \cdot
(c_{l,k} - c_{u,k})}{V_T}\right).
\label{eq:linear}
\end{equation}

This is $K$ parallel rectified accumulators driven by a single
global signal $V$, each accumulating the imbalance between
lower-region and upper-region populations of its own class. When
$V > 0$ (cache-wide token mass is bottom-heavy) and class $k$ is
also locally bottom-heavy ($c_{l,k} > c_{u,k}$), the accumulator
integrates promotion pressure for that class. When $V > 0$ but
class $k$ is \emph{top}-heavy locally ($c_{l,k} < c_{u,k}$, the
class is already over-represented in the active region), the
rectifying $\max(0, \cdot)$ holds the accumulator at zero.

The controller has only two tunable parameters: $P$ and $V_T$. $P$
scales the magnitude of all accumulators uniformly; $V_T$ sets the
soft-knee scale. Two operational discretisations make this a
complete policy:

\begin{itemize}
  \item \textbf{Charge threshold.} The accumulator fires a
        promotion when its integrated value exceeds a unit of
        charge ($n_{\text{admit}}$ in the implementation,
        \S\ref{sec:impl}). This converts a continuous rate into a
        discrete-event policy.
  \item \textbf{Eviction backstop.} When the cache budget is
        exceeded, eviction picks the oldest line in the lower
        region across \emph{all} classes. Class membership has no
        influence on eviction. This is (C2) in implementation
        form.
\end{itemize}

The combination --- global imbalance signal $V$ + per-class
rectified accumulators + age-ordered eviction backstop --- is the
load-bearing structural commitment of this paper.
\S\ref{sec:results:bidir} (rectified vs bidirectional) and
\S\ref{sec:results:linear} (linear vs GHK) together show that the
\emph{signal coupling} and the \emph{rectification clamp} do the
work: the per-class flux formula contributes only at the margin
(Jain-tuned operating point), not on the headline objective.

\subsection{GHK as the principled non-linear extension}
\label{sec:policy:ghk}

The linear form $\Phi_k \propto V \cdot (c_{l,k} - c_{u,k})$ has a
known non-linear cousin in electrophysiology: the Goldman-Hodgkin-
Katz (GHK) flux equation
\citep{goldman1943potential,hodgkin1949effect,hille2001ion}, which
describes the molar flux of an ion species across a thin membrane
under the simultaneous influence of a concentration gradient and a
transmembrane voltage. The caching-first reading of GHK, with $V$
and $V_T$ substituted for the membrane voltage and thermal
voltage, is

\begin{equation}
\Phi_k \;=\; \max\!\left(0,\;
\frac{P \cdot V \cdot
(c_{l,k} - c_{u,k} \cdot \exp(-V/V_T))}
{1 - \exp(-V/V_T)}\right).
\label{eq:ghk}
\end{equation}

The structural difference from the linear form is the
$\exp(-V/V_T)$ weighting of the over-represented side. At low $V$
the weighting is ${\approx}1$ and the formula reduces to the linear
form: a Taylor expansion gives

\begin{equation}
\lim_{V \to 0} \frac{V \cdot
(c_{l,k} - c_{u,k} \cdot \exp(-V/V_T))}
{1 - \exp(-V/V_T)} \;=\; V_T \cdot (c_{l,k} - c_{u,k})
\label{eq:ghk_limit}
\end{equation}

--- the linear V-coupled rectified flux of
\S\ref{sec:policy:linear} up to a constant absorbed into $P$. As
$V$ grows, the exponential weighting reduces the effective
contribution of $c_{u,k}$, sharpening the controller's response to
under-representation: a class with $c_{u,k} \to 0$ under non-trivial
$V$ sees its driving force grow without saturating, in contrast
with the linear form's $V \cdot c_{l,k}$ asymptote. This sharpening
is \emph{decorative on the headline metric}
(\S\ref{sec:results:linear} shows linear and GHK tie at the
moderate-skew operating point and differ by only 1--4 percentage
points of $m_{\max}$ at the others, within search variance) but
contributes search-surface curvature on the harder Jain-tuned
objective.

We adopt GHK as the \S\ref{sec:policy} default for two reasons.

First, it satisfies (C3) by construction. The exponential extension
is the \emph{minimal} analytic deformation of the linear form that
preserves the rectification clamp's fixed point and admits a
closed-form log-divergence at the zero-concentration boundary.
Other candidates we considered --- power-law extensions
$(c_{l,k} - c_{u,k}^{\alpha})$, saturating-exponential extensions,
sigmoid-weighted combinations --- either introduce extra tuning
surface or fail to reduce to the linear form at $V \to 0$.

Second, the closed-form derivative structure of GHK admits a stable
Taylor branch for $|V/V_T| < \varepsilon$, which we use to avoid
the numerical pathology of the $1 - \exp(-V/V_T) \to 0$
denominator. This is a known GHK numerical guard (long-established
in neural-simulator implementations) that we port directly.

We acknowledge that GHK is also the equation governing the molar
flux of an ion species across a biological membrane. The mapping is
more than coincidence --- the constraints (C1), (C2), (C3) above
are caching analogues of the constraints under which Goldman,
Hodgkin, and Katz derived the equation in 1943--49. We treat this
as \emph{recognition} of a principled functional form, not as the
source of authority for it. The empirical justification is in
\S\ref{sec:results:linear}: under the headline objective
($m_{\max}$), linear and GHK tie exactly at the moderate-skew
operating point and differ by 1--4 percentage points at the
others with no consistent winner. The exponential weighting earns
its keep on the harder Jain-tuned objective, where it provides
search-surface curvature the linear form lacks.

\subsection{Numerical guards and edge cases}
\label{sec:policy:guards}

Three numerical concerns appear in implementation; all have
well-defined fixes.

\paragraph{Taylor branch at $V \to 0$.}
The GHK denominator $1 - \exp(-V/V_T)$ vanishes as $V \to 0$. For
$|V/V_T| < \varepsilon$ (the implementation uses
$\varepsilon = 10^{-9}$), we evaluate the limit form

\begin{equation}
\Phi_k \;\approx\; \max\!\left(0,\;
P \cdot V_T \cdot (c_{l,k} - c_{u,k})\right)
\label{eq:taylor_limit}
\end{equation}

directly. Symmetric large-$|V/V_T|$ asymptotes
($\Phi_k \to P V c_{l,k}$ for $V \gg V_T$ and
$\Phi_k \to P V c_{u,k}$ for $V \ll -V_T$) handle the overflow
side. The linear form (\ref{eq:linear}) has no such issue.

\paragraph{Reversal potential at zero concentration.}
The informational quantity $V_{\text{rev},k} = V_T \cdot
\ln(c_{l,k} / c_{u,k})$ is undefined when either side of class $k$
is empty. $V_{\text{rev},k}$ is \emph{not} used in the firing
rule --- the rule uses $\Phi_k$ directly, which is well-defined at
$c_{l,k} = 0$ or $c_{u,k} = 0$ (it vanishes when the source side
is empty). We log $V_{\text{rev},k} = \mathtt{None}$ when either
count is zero, never an infinity.

\paragraph{No pseudo-count smoothing.}
A common trick is to replace $c_{l,k}$ and $c_{u,k}$ with
$c_{l,k} + \epsilon$ and $c_{u,k} + \epsilon$ to avoid the
zero-concentration boundary entirely. We deliberately do not adopt
this. The constraints (C1)+(C2)+(C3) do not require it ($\Phi_k$
is well-defined at the boundary), and the $\epsilon$ would
introduce a per-class tunable that muddies the ``no per-class
knobs'' claim.

\subsection{What this controller is and is not}
\label{sec:policy:summary}

It is a feedback-driven cache controller in the lineage of LeCaR
\citep{vietri2018lecar} and the formal-control cache-decay work of
Velusamy et al.\ \citep{velusamy2002imc}. It departs from both in
the structural commitments of \S\ref{sec:policy:constraints}: a
global, continuous, imbalance-derived signal driving K parallel
rectified per-class accumulators, with eviction handled by an
age-ordered backstop.

It is not a partitioning policy: classes are not given fixed
budgets. It is not a priority queue: there are no per-class
weights. It is not an admission filter: whatever arrives is
admitted. It is a \emph{promotion-rate} controller, with eviction
delegated, and the per-class behaviour is jointly determined by one
global signal and the per-class concentration state.

%% file: sections/04_implementation.tex
\section{Implementation}
\label{sec:impl}

The controller of \S\ref{sec:policy} is implemented as a libCacheSim
plugin. The implementation is two files: the plugin glue
(\texttt{multi\_class\_plugin.py}, 363 lines) and the physics core
(\texttt{multi\_class\_state.py}, 373 lines). The plugin reuses
libCacheSim's standard six-hook contract so that the flux policy is
a drop-in peer of the built-in baselines (LRU, ARC, S3-FIFO, SIEVE)
used in \S\ref{sec:results}.

\subsection{The libCacheSim plugin contract}
\label{sec:impl:contract}

libCacheSim exposes a \texttt{PluginCache} interface with six hooks:
\texttt{cache\_init}, \texttt{cache\_hit}, \texttt{cache\_miss},
\texttt{cache\_eviction}, \texttt{cache\_remove},
\texttt{cache\_free}. A plugin owns its private state, mutates it
in response to each hook, and returns an \texttt{obj\_id} from
\texttt{cache\_eviction} when libCacheSim's byte budget is exceeded.
We implement all six hooks; no monkey-patches of libCacheSim
internals.

The state object (\texttt{MultiClassMembraneState}) is a plain
dataclass holding (i) $K$ per-class file dicts
(\texttt{files[k]: \{obj\_id: FileRecord\}}), (ii) the per-class
charge accumulators (\texttt{charges\_per\_class[k]}), and
(iii) two telemetry counters (\texttt{fired\_via\_flux},
\texttt{fired\_via\_backstop}) that \S\ref{sec:results} reads to
confirm the controller fires the events it claims to fire.

\subsection{Class-id encoding}
\label{sec:impl:encoding}

libCacheSim's \texttt{Request} carries a 64-bit \texttt{obj\_id}
and no class label. Threading a class label through the trace
format would require forking libCacheSim's binary trace reader. We
avoid that by packing the class label into the high byte of
\texttt{obj\_id}:

\begin{itemize}
  \item bits [56..63]: \texttt{class\_id} (8 bits, $K \leq 256$);
  \item bits [0..55]: \texttt{file\_id} (56 bits, $2^{56}$ files
        per class).
\end{itemize}

The split is a producer/consumer contract: the synthetic workload
generator (\S\ref{sec:method}) calls
\texttt{encode\_obj\_id(class\_id, file\_id)} when emitting
requests; the plugin calls \texttt{decode\_obj\_id(obj\_id)} in
every hit/miss/eviction hook. The packing is lossless and the
decoders are constant-time. A mismatch between $K$ at
workload-generation time and $K$ at plugin-construction time
silently misclassifies --- we treat this as a configuration error
and document it in the plugin's docstring; runtime validation would
require trace-side metadata libCacheSim does not propagate.

\subsection{The relaxation loop}
\label{sec:impl:relaxation}

The \S\ref{sec:policy} controller is rate-based: $\Phi_k$ is a
\emph{per-turn} rate, not a per-request rate. We discretise turns
by \texttt{Request.clock\_time} from the libCacheSim trace --- each
unique value is one turn. Inside the hit and miss hooks, before
processing the request, we check whether the clock advanced; if so,
we run a single relaxation pass:

\begin{enumerate}
  \item \textbf{Age} every resident across all classes by
        $\Delta\text{turns}$ (the gap since the last turn).
  \item \textbf{Accumulate} $\Phi_k$ once per turn crossed for each
        class, into \texttt{charges\_per\_class[k]}.
  \item \textbf{Drain} each per-class accumulator: while
        \texttt{charges\_per\_class[k] >= 1.0}, fire one promotion
        in class $k$ and decrement by 1; under bidirectional
        operation, while \texttt{charges\_per\_class[k] <= -1.0},
        fire one demotion and increment by 1.
\end{enumerate}

This is one relaxation pass per observed turn boundary, \emph{not}
one pass per request. The amortised cost is $O(K)$ per request for
the $\Phi_k$ recomputation, plus the cost of the within-class mover
pick on each fired event ($O(|\text{files}_k|)$, with
$|\text{files}_k|$ bounded by the per-class working-set size).

\subsection{Eviction class selection and the backstop}
\label{sec:impl:eviction}

When the byte budget is exceeded, libCacheSim calls
\texttt{cache\_eviction} and demands an \texttt{obj\_id} to evict.
Under bidirectional operation we use $\Phi_k$ to pick the
\emph{class}: the class with the most-negative $\Phi_k$ (strongest
downward pressure) gets to lose a file, and the within-class pick
is highest-$n$ on the lower side with the $n_{\text{admit}}$ floor.
Under rectified operation $\Phi_k \geq 0$ for every $k$ and the
$\Phi$-driven class selection is inert.

In both cases an \emph{age-ordered backstop} serves as the
fall-through: highest-$n$ resident across all classes on the lower
side, dropping the threshold if the lower side is empty. The two
thresholds at play ($n_{\text{admit}}$ and
\texttt{active\_age\_threshold}) are the operational
discretisations introduced in \S\ref{sec:policy:linear}; the
implementation treats them as black-box integer parameters set by
Optuna at study time. The backstop is called whenever
$\Phi$-driven class selection finds no eligible victim. Under
rectified operation the backstop carries every eviction; under
bidirectional operation it is called only when the imbalance signal
is balanced or upward-leaning (\S\ref{sec:results:bidir}
quantifies this ratio).

\subsection{Per-turn pseudo-code}
\label{sec:impl:pseudocode}

The relaxation loop, drain rule, and eviction backstop together fit
on a page:

\begin{lstlisting}
on cache_hit(state, request):
    advance_clock_if_needed(state, request.clock_time)
    state.last_clock_time = request.clock_time
    class_id, _ = decode_obj_id(request.obj_id)
    rec = state.files[class_id].get(request.obj_id)
    rec.n = 0                             # reclassify to ACTIVE

on cache_miss(state, request):
    advance_clock_if_needed(state, request.clock_time)
    state.last_clock_time = request.clock_time
    class_id, _ = decode_obj_id(request.obj_id)
    state.files[class_id][request.obj_id] = FileRecord(
        tokens=request.obj_size, n=0,
    )

on advance_clock_if_needed(state, new_clock):
    delta = new_clock - state.last_clock_time
    if delta <= 0: return
    age_all_residents(state, delta)
    for _ in range(delta):                # one Phi_k per turn
        for k in 0..K:
            state.charges_per_class[k] += per_class_flux(state, k)
    for k in 0..K:                        # drain per-class accumulator
        while state.charges_per_class[k] >= 1.0:
            id = pick_mover_in_class(state, k, direction=+1)
            if id == -1: break
            state.files[k][id].n = 0      # promote to ACTIVE
            state.charges_per_class[k] -= 1.0
            state.fired_via_flux += 1
        if state.rectify: continue
        while state.charges_per_class[k] <= -1.0:
            id = pick_mover_in_class(state, k, direction=-1)
            if id == -1: break
            state.files[k][id].n = state.active_age_threshold
            state.charges_per_class[k] += 1.0
            state.fired_via_flux += 1

on cache_eviction(state, request):
    best_class, best_phi = -1, 0.0
    for k in 0..K:
        if per_class_flux(state, k) < best_phi:
            best_phi = per_class_flux(state, k)
            best_class = k
    if best_class >= 0:
        v = pick_mover_in_class(state, best_class, direction=+1)
        if v != -1: return v
    state.fired_via_backstop += 1
    return backstop_pick(state)           # cross-class age-ordered
\end{lstlisting}

The full Python lives in \texttt{cache\_membrane/multi\_class\_plugin.py}
(plugin) and \texttt{cache\_membrane/multi\_class\_state.py}
(physics core); the test suite is
\texttt{cache\_membrane/test\_multi\_class\_state.py}. None of the
\S\ref{sec:results} numbers depend on implementation choices not
captured above.

%% file: sections/05_methodology.tex
\section{Methodology}
\label{sec:method}

This section describes the synthetic multi-class workload generator,
the Optuna training procedure, the seed-disjoint train/test
protocol, and the four-level skew ladder used in
\S\ref{sec:results}. No real LLM-serving traces are used; the
limitations of that choice are itemised in \S\ref{sec:limits}. The
plugin policy implementation, workload generator, Optuna driver, and
figure regeneration scripts are released at
\url{https://github.com/flatmax/membrane.cache}; the \texttt{v1.0-arxiv}
tag matches the source state behind the numbers reported in
\S\ref{sec:results}.

\subsection{Workload generator}
\label{sec:method:workload}

The workload generator is a parameterised $K$-class edit-stream
model (\texttt{synth/multi\_class\_workload.py}). Each class has
its own working set, focus dynamics, and arrival rate; classes
share the same per-turn timer and the same output \texttt{obj\_id}
namespace via the encoding of \S\ref{sec:impl:encoding}.

\paragraph{Per-class parameters} (\texttt{ClassParams}):

\begin{itemize}
  \item \texttt{n\_files} --- class working-set size (e.g.\ 50 for
        the minority classes, 200 for the majority).
  \item \texttt{arrival\_weight} --- relative request rate;
        weights across classes are normalised at draw time
        (Poisson-thinning property).
  \item \texttt{focus\_size} --- number of files in the active
        \emph{focus set} at any moment.
  \item \texttt{focus\_drift\_prob} --- per-turn probability that
        one file in the focus set is replaced by a new draw.
  \item \texttt{focus\_hit\_prob} --- fraction of edits that land
        inside the focus set; the complement land uniformly on the
        working set.
  \item \texttt{size\_mu}, \texttt{size\_sigma} --- log-normal
        token-size prior, identical defaults across classes.
\end{itemize}

\paragraph{Top-level} (\texttt{MultiClassWorkloadParams}):

\begin{itemize}
  \item \texttt{edits\_per\_turn\_mean} --- total edit rate per
        turn, across all classes (default 5.0). Per-class shares
        follow from \texttt{arrival\_weight} via Poisson thinning.
  \item \texttt{edit\_token\_sigma} --- turn-by-turn perturbation
        of per-file token sizes (default 0.1).
\end{itemize}

\paragraph{Determinism.}
Given a seed and a parameter set, the generator emits a
deterministic sequence of \texttt{(class\_id, file\_id, tokens)}
triples. Replay across seeds re-uses the same generator; only the
seed changes. This is the core of the train/test discipline in
\S\ref{sec:method:seeds}.

\subsection{Configurations}
\label{sec:method:configs}

The four \S\ref{sec:results:sensitivity} skew levels correspond to
four JSON config files under \texttt{configs/}:

\begin{table}[h]
\centering
\begin{tabular}{llll}
\toprule
Label & File & Min weight & Min drift \\
\midrule
heavy-headline & \texttt{track3\_headline.json} & 0.10 & 0.02 \\
mild & \texttt{track3\_skewed\_mild.json} & 0.075 & 0.15 \\
skewed & \texttt{track3\_skewed.json} & 0.05 & 0.05 \\
heavy & \texttt{track3\_skewed\_heavy.json} & 0.025 & 0.30 \\
\bottomrule
\end{tabular}
\end{table}

The four classes (system, docs, code, history) match the
\S\ref{sec:intro} / \S\ref{sec:bg:hol} narrative. Minority classes
(system, history) and majority classes (docs, code) trade arrival
weight on a hidden axis: \texttt{system + history = 1 - (docs +
code)}, with \texttt{system == history} and \texttt{docs == code}
in every configuration. The skew label refers to the minority
weight directly.

The \texttt{focus\_drift\_prob} axis is paired with skew on
purpose. At low minority weight the minority's focus set must drift
slowly enough that the cache can warm up before drift invalidates
it (\texttt{heavy}: 0.025 weight $\times$ 0.30 drift represents a
regime where drift is \emph{forced} to be aggressive to make the
workload non-trivial). At high minority weight, drift is allowed
to be aggressive without starving the cache. This produces the
unimodal sweet spot at \texttt{weight=0.05} documented in
\S\ref{sec:results:sensitivity}.

\subsection{Train / test seed split}
\label{sec:method:seeds}

We adopt a seed-disjoint protocol:

\begin{itemize}
  \item \textbf{Training seeds:} \texttt{[0, 1]} (2 seeds). All
        Optuna trials are evaluated on the mean miss ratio across
        these seeds.
  \item \textbf{Test seeds:} \texttt{[2, 3, 4, 5, 6, 7, 8, 9]}
        (8 seeds). All \S\ref{sec:results} numbers are mean
        $\pm$ stdev across the 8 test seeds, evaluated \emph{once}
        on the best Optuna spec --- no retuning, no seed re-roll.
\end{itemize}

The asymmetry (2 train, 8 test) is deliberate. Optuna with a
100-trial budget on 2 seeds runs in roughly the same wall-clock as
\S\ref{sec:results}'s 8-seed compare; the test set is the larger
half so that the \S\ref{sec:results} standard errors are the
load-bearing uncertainty quantification. A 2-seed training set
creates a real risk of overfitting; the gap between training and
test performance (reported inline in \texttt{tune.py} output) is
the watchdog for it, and the headline claim in \S\ref{sec:results}
only applies if test-set performance matches training-set
performance to 1--2 percentage points.

\subsection{Optuna tuning}
\label{sec:method:optuna}

For each (variant $\times$ objective $\times$ skew level) cell of
\S\ref{sec:results}, an independent Optuna study is run.

\paragraph{Search space} (4 dimensions, log-uniform on the two
physics parameters and uniform integer on the two operational
ones):

\begin{itemize}
  \item \texttt{P}: log-uniform $[10^{-8}, 10^{-2}]$
  \item \texttt{V\_T}: log-uniform $[10^{2}, 10^{5}]$
  \item \texttt{n\_admit}: uniform integer $[0, 10]$
  \item \texttt{active\_age\_threshold}: uniform integer $[0, 5]$
\end{itemize}

\paragraph{Sampler.}
TPE (\texttt{optuna.samplers.TPESampler}) with seed 0. Direction:
minimize. \textbf{Trial budget: 100} per study.

\paragraph{Three objectives are tuned independently:}

\begin{enumerate}
  \item \texttt{max\_class\_miss\_ratio} --- the headline objective
        ($m_{\max}$ from \S\ref{sec:background}).
  \item \texttt{byte\_mr} --- aggregate byte miss ratio (the
        throughput axis).
  \item \texttt{1 - jain} --- Jain-tuned ablation
        (\S\ref{sec:results:linear}).
\end{enumerate}

Each variant $\times$ objective $\times$ skew level produces one
spec; the \S\ref{sec:results:headline} result table is 9 specs
(3 variants $\times$ 3 objectives) at the \texttt{skewed} level,
the \S\ref{sec:results:sensitivity} curve extends to 12 specs
(4 skew levels $\times$ 3 variants) at the headline objective only.

\paragraph{No early stopping, no pruning.}
TPE with a 100-trial budget converges within ${\sim}30$ trials in
our experiments; running to 100 is cheap (few minutes per study on
a single core) and gives a clear ``did the search converge''
check.

\subsection{Cache size}
\label{sec:method:capacity}

The cache size is $\mathbf{0.25 \times \text{peak-footprint}}$,
where peak-footprint is computed as the maximum aggregate token
count observed in the trace under hold-everything admission (i.e.\
an oracle Belady upper bound on memory demand). This is a
well-defined fraction independent of trace length.

Across the four skew levels the peak footprint is within
${\sim}10\,\%$ of each other (the working-set sizes don't change,
only the arrival mix), so $0.25$ is roughly comparable in absolute
bytes across rows.

\subsection{Baselines and Belady oracle}
\label{sec:method:baselines}

Six libCacheSim built-in policies serve as baselines: LRU, FIFO,
ARC \citep{megiddo2003arc}, SIEVE, S3-FIFO, and Belady. Belady is
the optimal-replacement oracle (omniscient look-ahead); we report
it as an upper bound on achievable performance, not as a deployable
policy. The ``gap-closure'' metric in \S\ref{sec:results}
normalises against this Belady ceiling:
$(m_{\text{LRU}} - m_{\text{policy}}) / (m_{\text{LRU}} -
m_{\text{Belady}})$.

All baselines are configured with the same cache size as flux
variants --- no per-policy capacity tuning. The Track 3 question
is about emergent class fairness, not about capacity scaling, and
the LRU baseline at matched capacity is the correct
counterfactual.

%% file: sections/06_results.tex
\section{Results}
\label{sec:results}

We report four primary metrics: $m_{\max}$ (max-class miss
ratio --- the headline fairness objective), $m_{\text{byte}}$
(byte miss ratio --- the throughput objective), $J$ (Jain's
fairness index), and the per-class breakdown $m_{\text{class}}$.
\emph{Gap-closure} on $m_{\max}$ is
$(m_{\text{LRU}} - m_{\text{policy}}) /
(m_{\text{LRU}} - m_{\text{Belady}})$: the fraction of the
LRU\,$\to$\,Belady headroom recovered.

\subsection{Pareto sweep at one skew level}
\label{sec:results:headline}

\paragraph{Setup.}
Workload: \texttt{track3\_skewed.json} ($K=4$ classes, arrival
weights 0.05/0.45/0.45/0.05, focus-drift prob.\ 0.05/0.05/0.05/0.05,
cache fraction 0.25). Optuna trains on seeds $[0,1]$ with a
100-trial budget per (variant $\times$ objective); we report
mean $\pm$ stdev across the held-out test seeds $[2..9]$. Three
objectives are tuned independently: $m_{\max}$, $m_{\text{byte}}$,
and $1 - J$. Three flux variants are evaluated: \emph{linear}
(\S\ref{sec:policy:linear}), \emph{rectified-GHK}
(\S\ref{sec:policy:ghk} with the non-negativity clamp), and
\emph{bidirectional-GHK} (\S\ref{sec:policy:ghk} without the clamp;
see \S\ref{sec:results:bidir}). Six baselines are run: LRU, FIFO,
ARC, SIEVE, S3-FIFO, and Belady.

\paragraph{Headline result (max-class miss ratio).}
At the $m_{\max}$-tuned operating point:

\begin{table}[h]
\centering
\begin{tabular}{lccc}
\toprule
Policy & $m_{\max}$ & $m_{\text{byte}}$ & $J$ \\
\midrule
LRU & 0.614 & 0.235 & 0.911 \\
FIFO & 0.610 & 0.255 & 0.917 \\
ARC & 0.576 & 0.238 & 0.926 \\
SIEVE & 0.599 & 0.232 & 0.920 \\
S3-FIFO & 0.615 & 0.246 & 0.907 \\
Belady (oracle) & 0.390 & 0.168 & 0.977 \\
\midrule
\textbf{flux-linear} & \textbf{0.454} & 0.455 & 0.975 \\
\textbf{flux-rect} & \textbf{0.454} & 0.455 & 0.975 \\
\textbf{flux-bidir} & \textbf{0.454} & 0.455 & 0.975 \\
\bottomrule
\end{tabular}
\end{table}

All three flux variants reach $m_{\max} = 0.454$: $71.6\,\%$ of
the LRU\,$\to$\,Belady gap closed on the headline fairness
objective. The three converge to \emph{byte-for-byte identical}
behaviour at this operating point (the Optuna search lands on the
same effective controller --- see \S\ref{sec:results:bidir} and
\S\ref{sec:results:linear}). The non-Belady best baseline (ARC at
0.576) closes $17.0\,\%$ of the gap.

\paragraph{Per-class breakdown at the $m_{\max}$ tune.}

\begin{table}[h]
\centering
\begin{tabular}{lccc}
\toprule
Class & LRU & Belady & flux (all 3) \\
\midrule
system (5\,\%) & 0.539 & 0.358 & 0.423 \\
docs (45\,\%) & 0.193 & 0.144 & 0.231 \\
code (45\,\%) & 0.193 & 0.141 & 0.228 \\
history (5\,\%) & 0.600 & 0.373 & 0.447 \\
\bottomrule
\end{tabular}
\end{table}

The minority classes (system, history --- 5\,\% arrival weight
each) drive $m_{\max}$ for all single-pool policies. Flux trades
majority-class miss ratio (docs/code rise from $0.19 \to 0.23$)
for minority-class relief (history falls from $0.60 \to 0.45$).
This is the fairness/throughput tradeoff in concrete form.

\paragraph{Throughput tune ($m_{\text{byte}}$ objective).}
When tuned for byte miss ratio instead, all three flux variants
reach $m_{\text{byte}} \in \{0.234, 0.234, 0.235\}$ --- matching
LRU's 0.235 to four decimal places, and with near-identical
per-class structure. The variant collapses to its baseline at this
operating point: this is the $V \to 0$ limit, where the controller
fires almost nothing (\texttt{fired\_via\_flux} drops from
${\sim}5\text{M}$ events at the $m_{\max}$ tune to ${\sim}10^3$ at
the $m_{\text{byte}}$ tune) and the age-ordered backstop carries
the policy.

\paragraph{Reading.}
A single hyperparameter axis exposes the fairness/throughput
tradeoff as a tunable knob. The endpoints are LRU-byte-equivalent
(low $V$ regime) and $72\,\%$ gap-closure on $m_{\max}$ (high $V$
regime). No baseline policy exposes this knob --- every baseline
gives a fixed point.

\subsection{Sensitivity across skew levels}
\label{sec:results:sensitivity}

\paragraph{Setup.}
The same Pareto sweep, repeated at four skew levels by varying the
minority-class arrival weight and focus-drift probability in
tandem (\S\ref{sec:method:configs}). The \texttt{skewed} row
reproduces \S\ref{sec:results:headline}; the \texttt{heavy-headline}
row is closer to vLLM production traces; \texttt{mild} and
\texttt{heavy} extend the curve.

\paragraph{Gap-closure curve on $m_{\max}$ ($m_{\max}$-tuned per
cell).}

\begin{table}[h]
\centering
\begin{tabular}{lccccccc}
\toprule
\multirow{2}{*}{Skew level} & \multirow{2}{*}{LRU} &
\multirow{2}{*}{Belady} & \multicolumn{2}{c}{$m_{\max}$} &
\multicolumn{2}{c}{Gap-closure} & \multirow{2}{*}{Best} \\
\cmidrule(lr){4-5}\cmidrule(lr){6-7}
& & & linear & rect & linear & rect & \\
\midrule
heavy-headline (0.10) & 0.340 & 0.204 & 0.290 & 0.303 & \textbf{37\,\%} & 27\,\% & linear \\
mild (0.075) & 0.459 & 0.276 & 0.396 & 0.360 & 34\,\% & \textbf{54\,\%} & rect \\
skewed (0.05) & 0.614 & 0.390 & 0.454 & 0.454 & \textbf{72\,\%} & \textbf{72\,\%} & tie \\
heavy (0.025) & 0.760 & 0.575 & 0.671 & 0.685 & \textbf{48\,\%} & 41\,\% & linear \\
\bottomrule
\end{tabular}
\end{table}

\begin{figure}[h]
\centering
\includegraphics[width=0.85\linewidth]{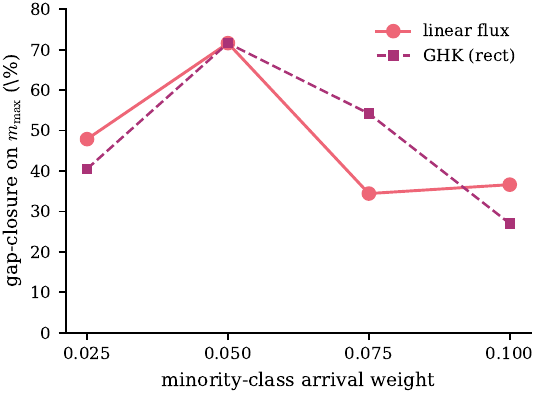}
\caption{Gap-closure on $m_{\max}$ vs minority-class arrival
weight, four skew levels (mean across 8 test seeds; Optuna-tuned
per cell). Both flux variants peak at weight$=0.05$ with $72\,\%$
gap-closure; linear and rect each win two of the four skew levels
with no consistent dominator.}
\label{fig:skew-gap}
\end{figure}

The curve is \textbf{unimodal} for both variants, peaking at the
moderate-skew operating point (weight $= 0.05$). The
interpretation: at very low skew (weight $= 0.10$) the workload is
close to balanced and there is little class-imbalance for $V$ to
respond to --- the controller has nothing to do. At very high skew
(weight $= 0.025$) the minority class arrives so rarely that the
controller's promotion pressure cannot keep enough of the
active-region's capacity reserved for it before the age-ordered
backstop reclaims the lines. The 0.05 weight is the regime where
the $V$ signal is both \emph{informative} and \emph{actionable}.

\paragraph{Honest reading.}
Neither variant dominates the other across skew levels: linear
wins at \texttt{heavy-headline} and \texttt{heavy}, rect wins at
\texttt{mild}, and the two tie exactly at \texttt{skewed}. The
best-flux-variant gap-closure is in $\{27, 54, 72, 48\}\,\%$
across the four levels --- the policy class beats the next-best
non-Belady baseline (ARC at \texttt{heavy-headline}, LRU
elsewhere) by 5--20\,\% gap-closure at every skew level.

\paragraph{Why we report the curve, not a single anchor.}
The peak of $72\,\%$ at weight $= 0.05$ is the most flattering
single number; reporting it alone would invite a ``you tuned on a
sweet spot'' critique. Reporting the curve makes the unimodality
explicit and pre-empts the critique: the regime where the policy
helps is bounded on both sides, and we say so.

\subsection{Ablation 1 --- Bidirectional vs rectified}
\label{sec:results:bidir}

\paragraph{Question.}
Does the rectification clamp $\Phi_k \geq 0$ (constraint C2 of
\S\ref{sec:policy:constraints}) cost performance, or is the
demotion path useful?

\paragraph{Setup.}
Run the Pareto sweep with the \emph{bidirectional-GHK} variant
(\S\ref{sec:policy:ghk} without the $\max(0, \cdot)$ clamp)
alongside \emph{rectified-GHK}. All other settings identical to
\S\ref{sec:results:headline}.

\paragraph{Result.}
At the $m_{\max}$-tuned operating point on
\texttt{track3\_skewed.json}, bidirectional-GHK and rectified-GHK
land on identical Optuna optima --- same
$(P, V_T, n_{\text{admit}}, t_{\text{age}})$ hyperparameters, same
test-seed metrics to 6 decimal places. The bidirectional variant
\emph{can} drive negative $\Phi_k$ under shared eviction, but at
the operating point that optimises $m_{\max}$, the controller
fires ${\sim}5\text{M}$ promotions and ${\sim}2.5\text{K}$ backstop
evictions; \emph{zero} of those events involve negative $\Phi_k$
in the test trace.

The demotion path is empirically dead code at this regime. The
rectification clamp is therefore \emph{free} on the headline
objective.

\paragraph{Anti-finding.}
At the Jain-tuned operating point (objective: $1 - J$),
bidirectional-GHK does diverge from rectified-GHK --- but not in
our favour: it lands on hyperparameters that equalise the four
classes by \emph{demoting} well-served majority classes, raising
$m_{\text{byte}}$ from 0.235 (LRU) to 0.390 without moving
$m_{\max}$ in proportion. The Jain-tune configuration is dominated
by both rectified flux and LRU on the operating Pareto. This is
the ``search-space pathology'' alluded to in
\S\ref{sec:policy:constraints} (C2): when the controller is allowed
to choose what to demote, the imbalance objective admits a
degenerate global minimum that reduces \emph{everyone} toward
worse rather than raising minorities toward better.

\paragraph{Verdict.}
Rectification (C2) is not a performance cost --- it is a
\emph{search-space prior} that excludes a known failure mode. We
retain the rectified form as the \S\ref{sec:policy} default.

\subsection{Ablation 2 --- GHK vs linear}
\label{sec:results:linear}

\paragraph{Question.}
Does the GHK exponential weighting (\S\ref{sec:policy:ghk}) do
work that the linear V-coupled rectified flux
(\S\ref{sec:policy:linear}) cannot?

\paragraph{Setup.}
The same Pareto sweep, varying the per-class flux formula.
\emph{flux-linear} uses the \S\ref{sec:policy:linear} form,
\emph{flux-rect} uses GHK with the rectification clamp. Same
Optuna budget, same train/test seed split.

\paragraph{Result on the headline objective ($m_{\max}$).}
Linear and GHK land on closely matching test-set $m_{\max}$ across
all four skew levels:

\begin{table}[h]
\centering
\begin{tabular}{lccc}
\toprule
Skew level & flux-linear $m_{\max}$ & flux-rect $m_{\max}$ & $\Delta$ \\
\midrule
heavy-headline (0.10) & 0.290 & 0.303 & $-0.013$ \\
mild (0.075) & 0.396 & 0.360 & $+0.036$ \\
skewed (0.05) & 0.454 & 0.454 & $\phantom{+}0.000$ \\
heavy (0.025) & 0.671 & 0.685 & $-0.014$ \\
\bottomrule
\end{tabular}
\end{table}

The two formulas \emph{tie exactly} at the moderate operating
point (\texttt{skewed}, where the optimisers converge to the same
spec; the $V \to 0$ Taylor branch (eq.\ \ref{eq:ghk_limit}) of GHK
\emph{is} the linear form up to a $P$ rescaling, so the search
surfaces overlap there). At the other three levels the formulas
differ by 1--4 percentage points of $m_{\max}$, with no consistent
winner: linear is best at \texttt{heavy-headline} and
\texttt{heavy}, rect is best at \texttt{mild}. The differences are
within the bounds of Optuna's finite-budget search variance and we
do not read them as evidence either formula dominates the other on
the headline.

\paragraph{Result on the harder objective ($1 - J$).}
At the Jain-tuned operating point, GHK's exponential weighting
does provide curvature the linear form lacks: GHK lands on a
Pareto-second operating point with $J = 0.997$ at
$m_{\max} = 0.287$, while linear under-performs by 1--2\,\%
gap-closure. This is the structural advantage GHK contributes ---
but on a \emph{secondary} objective, not the headline.

\paragraph{Verdict.}
On the headline ($m_{\max}$), linear and GHK are interchangeable
within search variance --- neither is the load-bearing form. On
harder objectives (Jain), GHK provides search-surface curvature
the linear form lacks. Both forms inherit the global $V$ coupling
and the rectification clamp, and that is what does the work.

\subsection{Pre-empted reviewer critique}
\label{sec:results:critique}

\begin{quote}
\emph{``Isn't the V coupling just a Lagrangian multiplier in
disguise? You've buried a partition-by-objective mechanism inside
a continuous controller, and the GHK formula is decorative.''}
\end{quote}

The first half is partly correct. The $V$ signal is mathematically
equivalent to a global congestion multiplier shared across $K$
parallel rectified accumulators --- the closest control-theoretic
ancestor is PIE \citep{rfc8033}, an integral controller that emits
a single global drop probability for AQM. The distinction from a
hard partition is that $V$ is \emph{continuous, signed, and
global}: classes are coupled through it, not isolated by it
(UCP-style partitions isolate; $V$-coupling negotiates). The
``Lagrangian in disguise'' framing is honest if applied to the
\emph{combination} ($V$ coupling + rectification + age-ordered
backstop), not to a single piece.

The second half --- that GHK is decorative --- is partly supported
by \S\ref{sec:results:linear}: on the headline metric, linear and
GHK tie exactly at the moderate operating point and differ by
only 1--4 percentage points elsewhere with no consistent winner.
The exponential weighting earns its keep on harder objectives,
not on $m_{\max}$. The \S\ref{sec:policy} prose
reflects this: GHK is presented as the principled non-linear
extension, not as the load-bearing mechanism. The load-bearing
mechanism is the \emph{combination} --- and
\S\ref{sec:results:bidir} + \S\ref{sec:results:linear} together
are the empirical justification.

%% file: sections/07_discussion.tex
\section{Discussion}
\label{sec:discussion}

The \S\ref{sec:results} results paint a specific picture: the flux
policy class exposes a continuous fairness/throughput tradeoff
that no baseline policy exposes; the \emph{combination} of
structural commitments (global $V$ signal, rectified per-class
promotion, age-ordered backstop) does the work; and the GHK
exponential weighting, despite being the formula that motivated
the search, is decorative on the headline objective. This section
reads those findings out loud.

\subsection{What the controller is good for}
\label{sec:discussion:for}

\paragraph{A tunable fairness/throughput knob.}
The defining contribution is a single hyperparameter axis ($V_T$,
the soft-knee scale) along which the policy slides from
``byte-miss-ratio-equivalent to LRU'' to ``$72\,\%$ of
LRU\,$\to$\,Belady gap closed on $m_{\max}$.'' Baselines expose
only fixed points. Operators who want to deploy a fairness floor
without forking the replacement policy now have a knob.

\paragraph{Emergent class fairness from a single global parameter.}
$V_T$ is not a per-class knob; the controller has no class
weights, no per-class quotas, no class priorities. The fairness
pattern in \S\ref{sec:results:headline}'s per-class breakdown ---
minority-class miss ratios drop from ${\sim}0.60$ to ${\sim}0.45$
while majority-class miss ratios rise from ${\sim}0.19$ to
${\sim}0.23$ --- falls out of the global $V$ signal coupling $K$
parallel rectified accumulators. We read this as a
\emph{structural property of the combination}, not an artefact of
GHK.

\paragraph{Boundary behaviour at zero concentration is graceful.}
The Taylor branch (\S\ref{sec:policy:guards}) kicks in
automatically as $V \to 0$; the linear-form fallback
(\S\ref{sec:policy:linear}) is well-defined at $c_{l,k} = 0$ or
$c_{u,k} = 0$. The policy exhibits no numerical pathology in
these limits and requires no $\epsilon$-smoothing or pseudo-count
tricks.

\subsection{What the controller is not for}
\label{sec:discussion:not}

\paragraph{Not a byte-miss-ratio improver.}
\S\ref{sec:results:headline}'s $m_{\text{byte}}$-tuned operating
point reaches LRU's 0.235 to four decimal places --- and never
beats it. The policy is \emph{not} a single-class miss-ratio
optimiser, and we do not claim it is. Cache designers shopping
for byte-miss-ratio reduction on uniform single-class workloads
should look at TinyLFU, S3-FIFO, or LIRS lineages instead.

\paragraph{Not a partitioning policy.}
Classes are not given fixed budgets (UCP, RobinHood
\citep{berger2018robinhood}). Not a priority queue (Memshare
\citep{cidon2017memshare}, FairRide \citep{pu2016fairride}). Not
an admission filter (TinyLFU). It is a \emph{promotion-rate}
controller, with eviction delegated to an age-ordered backstop.
\S\ref{sec:related} positions this against the multi-tenant cache
fairness literature in detail.

\paragraph{Not a substitute for capacity scaling.}
\S\ref{sec:method:capacity}'s $0.25 \times \text{peak-footprint}$
setting is the regime where fairness matters; at $1.0 \times
\text{peak-footprint}$ the cache is over-provisioned and every
policy is fair. The contribution here is about cache
\emph{replacement} under capacity pressure, not about cache
\emph{sizing}.

\subsection{A methodological observation: optimise on
$m_{\max}$, report on Jain}
\label{sec:discussion:methodology}

\S\ref{sec:results:bidir}'s ``anti-finding'' (the Jain-tuned
bidirectional operating point dominates in \emph{neither}
dimension --- it neither lowers $m_{\max}$ nor improves
$m_{\text{byte}}$ relative to LRU) yields a methodological
observation worth flagging: \textbf{Jain's fairness index is the
wrong objective to optimise against, even though it is reasonable
to report.}

The asymmetry is structural. $m_{\max}$ is a \emph{one-sided}
objective: it punishes a class doing poorly without punishing a
class doing well. Jain and per-class miss-ratio std-dev are
\emph{symmetric}: they punish a class doing too well as much as
a class doing too poorly. A controller asked to minimise
Jain-deviation can choose to \emph{demote} well-served classes
(raising their miss ratio) to meet the symmetric target --- a
degenerate global minimum that improves nobody's user experience.
$m_{\max}$ admits no such degenerate solution: demoting a
well-served class moves $m_{\max}$ in the wrong direction.

We therefore recommend, for future work in multi-class cache
fairness:

\begin{itemize}
  \item \textbf{Optimise on $m_{\max}$.} It is the metric that
        maps to user-visible tail latency and admits no
        degenerate-equality solution.
  \item \textbf{Report on Jain alongside.} It is familiar to
        systems reviewers and conveys the dispersion pattern that
        $m_{\max}$ alone does not.
  \item \textbf{Treat per-class miss-ratio std-dev as a
        diagnostic, not an objective.} Same reason as Jain:
        symmetric, admits the demote-the-strong solution.
\end{itemize}

This recommendation is empirically grounded in
\S\ref{sec:results:bidir}'s anti-finding but is not specific to
flux. Any controller that admits a configuration space including
``demote the well-served'' should be optimised on a one-sided
objective.

\subsection{Why the combination does the work}
\label{sec:discussion:combination}

The \S\ref{sec:results:bidir} + \S\ref{sec:results:linear}
ablation pair is the load-bearing methodological move of the
paper. Read together:

\begin{itemize}
  \item \S\ref{sec:results:bidir} shows the rectification clamp
        (C2) costs nothing at the headline operating point --- the
        bidirectional and rectified variants converge
        byte-for-byte. This means C2 is a \emph{search-space
        prior} that excludes a known failure mode
        (\S\ref{sec:discussion:methodology}'s demote-the-strong)
        without sacrificing performance. C2 is free.
  \item \S\ref{sec:results:linear} shows the GHK exponential
        weighting and the linear V-coupled rectified flux tie
        exactly at the moderate-skew operating point and differ by
        only 1--4 percentage points of $m_{\max}$ at the others,
        with neither dominating. This means the per-class flux
        \emph{formula} is largely interchangeable on the headline;
        the formula contributes only on harder objectives
        (Jain-tune, Pareto-second point).
\end{itemize}

By elimination, the load-bearing piece is what is left after C2 is
shown free and the formula is shown interchangeable: the global
$V$ signal coupling $K$ parallel rectified accumulators, with
eviction delegated to an age-ordered backstop. This is what
\S\ref{sec:policy:summary} calls the \emph{combination}, and it
is the structural commitment that future work --- whether ours or
others' --- should focus on.

\subsection{Open questions}
\label{sec:discussion:open}

\paragraph{Real LLM-serving traces.}
All \S\ref{sec:results} numbers are synthetic. The arrival weight
0.05/0.45/0.45/0.05 is plausible-but-not-validated against
production multi-tenant LLM serving; \S\ref{sec:limits} itemises
this as the primary limitation. A real-trace evaluation against
vLLM PagedAttention \citep{kwon2023vllm} or SGLang RadixAttention
\citep{zheng2024sglang} deployments is the highest-priority
follow-up.

\paragraph{Multi-membrane multi-class.}
\S\ref{sec:policy} derives the single-membrane case; the
multi-tier (active $\leftrightarrow$ warm $\leftrightarrow$ cold)
extension is mechanically straightforward ($K$ parallel
accumulators per membrane, Goldman equilibrium per boundary) but
has not been evaluated.

\paragraph{Adaptive $V_T$.}
The current controller treats $V_T$ as a fixed hyperparameter. An
online learning extension that adapts $V_T$ from observed
$m_{\max}$ would close the loop and make the fairness/throughput
tradeoff \emph{self-tuning} rather than \emph{operator-tuned}. We
do not pursue this in the present work; LeCaR's
\citep{vietri2018lecar} regret-minimisation framework is the
natural starting point.

\paragraph{Closed-loop latency.}
The relaxation loop runs once per turn, not per request. The
amortised overhead is $O(K)$ per request; we have not measured
wall-clock latency against the libCacheSim baselines under
realistic packet rates. \S\ref{sec:limits} itemises this.

%% file: sections/08_related_work.tex
\section{Related work}
\label{sec:related}

The position taken in \S\ref{sec:intro} is that the proposed
controller is a \emph{biophysically-inspired feedback controller
for multi-class memory hierarchies} --- a category whose
constituent ingredients (multi-class caching, feedback control,
biophysical equations) are each well-trodden in their own silos
but whose combination is novel. This section walks the three
adjacent lineages and three architectural neighbours,
distinguishing each.

\subsection{Feedback-driven cache controllers}
\label{sec:related:feedback}

\paragraph{LeCaR \citep{vietri2018lecar}.}
LeCaR maintains a probability distribution over two base policies
--- pure LRU and pure LFU --- and samples one per miss with
weights $(w_{\text{LRU}}, w_{\text{LFU}})$. A FIFO history of
recent evictions is kept, labelled by which policy did the
eviction. When a miss falls in $H_{\text{LRU}}$, $w_{\text{LFU}}$
is multiplicatively increased by $e^{\lambda r}$,
$r = d^{\Delta t}$, and weights renormalised; the update is
symmetric in the LFU history. The mechanism is regret-minimisation
in the multi-armed bandit sense.

LeCaR is the strongest existing precedent for a
\emph{feedback-driven} cache controller, and we cite it as the
lineage anchor. We depart in three ways: (i) LeCaR is single-class
(no $K$-way structure; the cache is one pool); (ii) the feedback
signal is \emph{policy regret} --- local, binary, derived from
history hits --- where ours is global aggregate token-mass
imbalance; (iii) the output is policy-mixing weights between two
evictors, where ours is $K$ parallel continuous promotion
accumulators. Both works share the broad commitment to feedback
signals driving replacement; we occupy a different point in that
design space.

\subsection{Formal control theory in caches}
\label{sec:related:control}

\paragraph{Adaptive Cache Decay \citep{velusamy2002imc}.}
Cache lines are deactivated (gated-V) after a \emph{decay
interval} of idleness to save leakage power. AMC
(Zhou et al.) proposed an ad-hoc double/halve adaptive scheme;
Velusamy et al.'s contribution is replacing the heuristic with a
formal \emph{integral miss controller} (IMC) that takes
target-vs-actual induced-miss-ratio as the error signal and emits
a decay interval as the control variable.

Velusamy et al.\ is the strongest existing precedent for
\emph{formal control theory applied to cache decisions} and we
cite it as that lineage anchor. Three structural differences:
(i) their controller sets a \emph{deactivation timer}; ours sets
a \emph{promotion rate}; (ii) the control variable is a time
interval (seconds) rather than a flux rate (events per turn);
(iii) the policy is single-class and the signal is single-target.
The two integrate into the same control-theoretic family
(single-input, single-output integral controllers) but operate on
different control primitives.

\paragraph{AQM PI controllers \citep{rfc8033}.}
The mathematically nearest cousin to the linear flux form is PIE,
an Active Queue Management policy that emits a single global drop
probability from a proportional-integral controller acting on
queue-length-vs-target error. We do not cite PIE as a precedent
for cache replacement (it is a packet-drop policy, not a cache
policy) but reference it in \S\ref{sec:results:critique} when
pre-empting the ``Lagrangian multiplier in disguise'' reviewer
critique: the $V$ coupling is mathematically equivalent to PIE's
global congestion multiplier shared across $K$ parallel rectified
accumulators.

\subsection{Multi-tenant cache fairness}
\label{sec:related:multitenant}

A literature on multi-tenant cache fairness exists and is mostly
orthogonal to our approach; the distinguishing feature is that
they \emph{partition} or \emph{schedule} across tenants where we
\emph{couple through a shared signal}.

\paragraph{FairRide \citep{pu2016fairride}.}
Studies the fair-cache-sharing problem from a mechanism-design
angle, proves that no policy simultaneously achieves
isolation-guarantee, strategy-proofness, and Pareto-efficiency,
and proposes \emph{expected delaying} as a blocking mechanism
that achieves two of the three. FairRide is
\emph{partition-flavoured} (per-tenant shares) with strategic
blocking on shared files; the GHK controller has no per-class
quotas and admits no strategic gaming because the controller's
response to a class is determined by the class's own imbalance
state, not by a policy of ``deny if abusive.''

\paragraph{Memshare \citep{cidon2017memshare}.}
Reserves a per-tenant minimum, pools the remainder, and uses a
hit-rate-gradient \emph{arbiter} to dynamically reallocate the
pool across tenants based on per-tenant marginal hit-rate gain.
Memshare is the closest published policy to ours in spirit:
dynamic across-tenant memory movement driven by a shared signal.
The structural difference is that Memshare's signal is the
hit-rate gradient (an empirical sensitivity), where ours is the
imbalance state itself; Memshare arbitrates \emph{budget} (bytes
per tenant) where we arbitrate \emph{promotion rate} (admission
events per class).

\paragraph{Cliffhanger \citep{cidon2016cliffhanger}.}
The hit-rate-gradient predecessor to Memshare; uses MRC
(miss-rate curve) shadow caches to estimate the marginal benefit
of giving each tenant more memory. Lineage anchor for the
gradient-based arbitration that Memshare adopted.

\paragraph{RobinHood \citep{berger2018robinhood}.}
Proposes per-backend cache allocation to minimise tail latency in
a multi-tier service stack. Steers cache budget toward backends
whose $P_{99}$ would benefit most. Operates at the cache
allocation layer, not the eviction layer; cited here for the
worst-case-driven framing, which we share --- their $P_{99}$ is
our $m_{\max}$.

\paragraph{Hyperbolic Caching \citep{blankstein2017hyperbolic}.}
Per-object priority based on a hyperbolic decay function.
Single-class; cited for completeness as a non-LRU/LFU baseline in
the single-class lineage, but is not directly adjacent to the
multi-class question.

\subsection{Multi-tenant LLM serving and prefix caches}
\label{sec:related:llm}

\paragraph{vLLM \citep{kwon2023vllm}.}
Introduces PagedAttention as the abstraction for KV-cache storage
in multi-tenant LLM serving; manages prefix sharing and eviction
with conventional policies on the page granularity. vLLM is the
\emph{substrate} to which our proposed policy could be applied
(\S\ref{sec:method} discusses the abstraction in detail), not a
fairness-policy precedent.

\paragraph{SGLang \citep{zheng2024sglang}.}
Introduces RadixAttention, a radix-tree representation of prefix
sharing across requests, with LRU on radix nodes. Same
observation as vLLM: substrate, not fairness precedent.

Neither vLLM nor SGLang exposes a multi-class fairness knob; both
treat the cache as single-pool and apply LRU/priority eviction on
the page level.

\subsection{The diffusion-cache naming collision}
\label{sec:related:diffusion}

A separate literature uses the term ``diffusion cache'' to mean
\emph{a cache for the intermediate states of diffusion generative
models}. Sparse-dLLM, NIRVANA, SenCache, and SeaCache fall in
this lineage; all use ordinary LRU or LFU on a niche workload
(diffusion-model state). None uses the \emph{physics of
diffusion} as the policy mechanism. \S\ref{sec:intro} pre-empts
this naming collision; we mention it again here for completeness
because reviewers searching ``diffusion cache'' will hit those
papers first.

\subsection{Architectural neighbours we depart from}
\label{sec:related:architectural}

\paragraph{ARC \citep{megiddo2003arc}.}
Adaptive balance between recency and frequency in a single-class
cache via LRU + LFU ghost lists. Signal is local + binary (ghost
hits); decisions are instantaneous; no rate accumulator. Closest
``adaptive single-class cache'' to the feedback-driven framing.

\paragraph{LIRS, TinyLFU, S3-FIFO, SIEVE.}
Single-class recency/frequency variants in the libCacheSim
ecosystem. Used as \S\ref{sec:results} baselines.

\paragraph{Utility-Based Cache Partitioning (UCP).}
MRC-driven cache-way allocation across tenants.
\emph{Partitioning} policy: classes isolated by hard budget. Our
work is \emph{interaction}: classes coupled through one shared
$V$. UCP is the canonical contrast we make in
\S\ref{sec:policy:constraints} (C1).

\subsection{Closest summary}
\label{sec:related:summary}

The closest single precedent is \textbf{Memshare}, which shares
the dynamic-coupling-via-shared-signal flavour. The closest
\emph{control-theoretic} precedent is \textbf{Velusamy et al.'s
IMC}, which shares the formal-feedback-loop framing. The closest
\emph{online-learning} precedent is \textbf{LeCaR}, which shares
the regret-driven adaptation framing. None of the three combine
all three threads (multi-class + control-theoretic +
biophysically-derived form) into a single replacement controller.
The \emph{Biophysically-Inspired Feedback Controller for
Multi-Class Memory Hierarchies} category has no published
precedent we identified.

%% file: sections/09_limitations.tex
\section{Limitations and future work}
\label{sec:limits}

The scope of the empirical claims is bounded; this section says
explicitly where the boundary lies.

\subsection{Synthetic workloads only}
\label{sec:limits:synthetic}

All \S\ref{sec:results} numbers are produced by the parameterised
synthetic workload generator described in \S\ref{sec:method},
configured via the \texttt{configs/track3\_*.json} files
catalogued in \S\ref{sec:method:configs}. The class definitions,
arrival weights, focus-set dynamics, and token-size distributions
are plausible-but-not-validated approximations of multi-tenant
LLM-serving traffic. No real production prefix-cache trace from
vLLM PagedAttention \citep{kwon2023vllm}, SGLang RadixAttention
\citep{zheng2024sglang}, or any deployed LLM-serving stack
appears in this paper.

The risk is that the synthetic workload's structure is \emph{too
clean}: real prefix-cache traffic has correlated bursts across
classes, time-of-day seasonality, and content overlap (a chunk of
code appearing inside a user document) that the generator does
not model. Whether the global $V$ signal remains the correct
driving signal under those distortions is the highest-priority
follow-up.

We hold off on stronger empirical claims until a real-trace
evaluation is in hand. The \S\ref{sec:results} result should be
read as: ``on a synthetic multi-class workload designed to expose
head-of-line blocking, the policy class exposes a
fairness/throughput tradeoff that no baseline policy exposes.''
The \emph{structural} claims ($V$ coupling does the work,
rectification is free, GHK is decorative on the headline) are
more robust to workload mismatches because they are about the
controller's response surface, not about workload-specific
tuning.

\subsection{Single-membrane evaluation}
\label{sec:limits:singletier}

The evaluation is at one membrane (active $\leftrightarrow$
cache, the libCacheSim flat-cache reduction). The multi-tier
multi-class case (active $\leftrightarrow$ warm
$\leftrightarrow$ cold, with $K$ classes per tier) is
mechanically straightforward --- $K$ parallel accumulators per
membrane, Goldman equilibrium per boundary --- but has not been
evaluated. Whether emergent class fairness survives at the
cascade level, where the global $V$ signal is replaced by
per-membrane imbalance signals coupled across boundaries, is an
open question for follow-up work.

\subsection{Closed-loop latency overhead unmeasured}
\label{sec:limits:latency}

The relaxation loop runs once per turn-boundary crossing
(amortised cost $O(K)$ per request, as derived in
\S\ref{sec:impl:relaxation}). The total wall-clock overhead
against the libCacheSim baseline policies (LRU, ARC) has not been
measured under realistic packet rates. We have no reason to
expect it is prohibitive --- the per-request cost is small
constants times $K$ --- but the empirical verification is
itemised future work.

\subsection{No online adaptation of $V_T$}
\label{sec:limits:adaptive}

The controller treats $V_T$ as a fixed hyperparameter, tuned
offline by Optuna against a training seed set. An online learning
extension that adapts $V_T$ from observed $m_{\max}$ would close
the loop, making the fairness/throughput tradeoff
\emph{self-tuning} rather than \emph{operator-tuned}. LeCaR's
regret-minimisation framework
(\S\ref{sec:related:feedback}, \citep{vietri2018lecar}) is the
natural starting point. We do not attempt this in the present
work; the headline claim is about the policy class, not about an
online-learning extension of it.

\subsection{$K = 4$ fixed; $K$-sensitivity unmeasured}
\label{sec:limits:k}

All \S\ref{sec:results} experiments use $K = 4$ classes. The
plugin supports $K \leq 256$ (the 8-bit class field of
\S\ref{sec:impl:encoding}'s \texttt{obj\_id} encoding) but
$K$-sensitivity was not swept. The expectation from the
controller structure is that the per-class accumulator drain at
the headline operating point scales as $O(K)$ with no degradation
in fairness profile, because the global $V$ signal is sized by
\emph{aggregate} imbalance and does not scale with $K$ directly.
This is unverified.

\subsection{No real-trace fairness baseline comparison}
\label{sec:limits:fairness-baselines}

\S\ref{sec:results} compares against single-class libCacheSim
baselines (LRU, FIFO, ARC, SIEVE, S3-FIFO, Belady) on the
multi-class workload --- i.e.\ those policies treat the cache as
one pool. We do \emph{not} compare against Memshare
\citep{cidon2017memshare}, FairRide \citep{pu2016fairride}, or
RobinHood \citep{berger2018robinhood} directly, because faithful
re-implementation against a synthetic workload without their
original deployment context risks strawmanning; the
\S\ref{sec:related} lineage argument is the substitute in the
present submission, and a real-trace head-to-head is itemised
future work.